\documentclass[compsoc,conference,a4paper,10pt,times]{IEEEtran}
\IEEEoverridecommandlockouts

\def\BibTeX{{\rm B\kern-.05em{\sc i\kern-.025em b}\kern-.08em
    T\kern-.1667em\lower.7ex\hbox{E}\kern-.125emX}}

\usepackage{cite} 
\usepackage{amsmath,amssymb,amsfonts}
\usepackage{algorithmic, algorithm}
\usepackage{graphicx}
\usepackage{textcomp}
\usepackage{xspace}
\usepackage{enumitem}
\usepackage{adjustbox,multirow}
\usepackage{caption}
\usepackage[labelfont=bf,skip=10pt, figurename=Fig.]{caption}
\usepackage{subcaption}
\usepackage{listings}
\usepackage[dvipsnames]{xcolor}
\usepackage{xcolor}
\usepackage[T1]{fontenc}
\usepackage{tikz}
\usepackage{filecontents}
\usepackage{booktabs}
\usepackage{float}
\usepackage{graphicx}
\usepackage{subfiles}
\usepackage{hyperref}
\usepackage[capitalize,nameinlink]{cleveref}
\usepackage{soul}
\usepackage{lipsum}
\usepackage{tabularx}

\renewcommand{\thetable}{\Roman{table}}

\renewcommand{\paragraph}[1]{\smallskip\noindent\textbf{#1.}}

\newcommand{\tool}{\textsc{CTINexus}\xspace}
\newcommand{\toolno}{\textsc{CTINexus}}
\newcommand{\kg}{CSKG\xspace}

\newcommand{\prefix}{cybersecurity\xspace}

\newcommand{\eat}[1]{}

\definecolor{light-gray}{gray}{0.95} 
\definecolor{comment-gray}{rgb}{0.41,0.6,0.33} 
\definecolor{strongred}{RGB}{190, 0, 0}

\lstdefinestyle{mystyle}{
    language=Python, 
    captionpos=b,  
    basicstyle=\fontfamily{phv}\selectfont\footnotesize,  
    keywordstyle=\bfseries,  
    breaklines=true, 
    breakatwhitespace=true, 
    showstringspaces=false, 
    numbers=left, 
    numberstyle=\tiny\color{black}, 
    numbersep=10pt, 
    commentstyle=\color{comment-gray},  
    backgroundcolor=\color{light-gray},  
    xleftmargin=17pt, 
    xrightmargin=3.4pt, 
    framextopmargin=15pt,  
    framexbottommargin=10pt,  
    framexleftmargin=17pt,  
    framesep=3pt, 
    fillcolor=\color{light-gray},  
    linewidth=\linewidth, 
    basewidth=0.5em,
    deletekeywords={[2]file},
    morekeywords={match, if, then, label_host, label_file, drop, alert, allow, declassify, endorse, contains}  
}

\lstdefinestyle{plaintext}{
    basicstyle=\ttfamily\footnotesize,  
    breaklines=true,                    
    breakatwhitespace=true,             
    showstringspaces=false,             
    frame=single,                       
    backgroundcolor=\color{light-gray}, 
    xleftmargin=17pt, 
    xrightmargin=3.4pt, 
    framextopmargin=5pt, 
    framexbottommargin=5pt, 
    framexleftmargin=17pt, 
    framesep=3pt, 
    fillcolor=\color{light-gray}, 
    linewidth=\linewidth,
    basewidth=0.5em,
}

\crefname{figure}{Fig.}{Figs.}
\crefname{listing}{Policy}{Policies}
\crefname{section}{Section}{Sections}
\crefname{table}{Table}{Tables}
\crefname{BNF}{Grammar}{Grammars}
\crefname{algorithm}{Algorithm}{Algorithms}

\hypersetup{%
	bookmarksnumbered, bookmarksopen=true, bookmarksopenlevel=1,%
      colorlinks = false,
}

\AtBeginEnvironment{appendices}{\crefalias{section}{appendix}\crefalias{subsection}{appendix}}

\newif\ifshowcomment
\showcommenttrue

\ifshowcomment
    \newcommand{\cici}[1] {{\footnotesize\color{Aquamarine}[CiCi: #1]}}
    \newcommand{\pgao}[1] {{\footnotesize\color{blue}[Peng: #1]}}
    \newcommand{\pgaoii}[1] {{\footnotesize\color{cyan}[Peng: #1]}}
    \newcommand{\osama}[1] {{\footnotesize\color{orange}[Osama: #1]}}
    \newcommand{\saimon}[1] {{\footnotesize\color{blue}[Saimon: #1]}}
    \newcommand{\dawn}[1] {{\footnotesize\color{blue}[Dawn: #1]}}
    
\else
    \newcommand{\cici}[1]{}
    \newcommand{\pgao}[1]{}
    \newcommand{\pgaoii}[1]{}
    \newcommand{\osama}[1]{}
    \newcommand{\saimon}[1]{}
    \newcommand{\dawn}[1]{}
    
\fi

\begin{document}

\title{\tool: Automatic Cyber Threat Intelligence Knowledge Graph Construction Using Large Language Models}

\author{\IEEEauthorblockN{Yutong Cheng}
\IEEEauthorblockA{\textit{Virginia Tech} \\
yutongcheng@vt.edu}
\and
\IEEEauthorblockN{Osama Bajaber}
\IEEEauthorblockA{\textit{Virginia Tech} \\
obajaber@vt.edu}
\and
\IEEEauthorblockN{Saimon Amanuel Tsegai}
\IEEEauthorblockA{\textit{Virginia Tech} \\
saimon.tsegai@vt.edu}
\and
\IEEEauthorblockN{Dawn Song}
\IEEEauthorblockA{\textit{UC Berkeley} \\
dawnsong@berkeley.edu}
\and
\IEEEauthorblockN{Peng Gao}
\IEEEauthorblockA{\textit{Virginia Tech} \\
penggao@vt.edu}
}

\maketitle

\begin{abstract}

Textual descriptions in cyber threat intelligence (CTI) reports, such as security articles and news, are rich sources of knowledge about cyber threats, crucial for organizations to stay informed about the rapidly evolving threat landscape. However, current CTI knowledge extraction methods lack flexibility and generalizability, often resulting in inaccurate and incomplete knowledge extraction. Syntax parsing relies on fixed rules and dictionaries, while model fine-tuning requires large annotated datasets, making both paradigms challenging to adapt to new threats and ontologies. To bridge the gap, we propose CTINexus, a novel framework leveraging optimized in-context learning (ICL) of large language models (LLMs) for data-efficient CTI knowledge extraction and high-quality cybersecurity knowledge graph (\kg) construction. Unlike existing methods, CTINexus requires neither extensive data nor parameter tuning and can adapt to various ontologies with minimal annotated examples. This is achieved through: (1) a carefully designed automatic prompt construction strategy with optimal demonstration retrieval for extracting a wide range of cybersecurity entities and relations; (2) a hierarchical entity alignment technique that canonicalizes the extracted knowledge and removes redundancy; (3) an long-distance relation prediction technique to further complete the \kg with missing links. Our extensive evaluations using 150 real-world CTI reports collected from 10 platforms demonstrate that CTINexus significantly outperforms existing methods in constructing accurate and complete \kg, highlighting its potential to transform CTI analysis with an efficient and adaptable solution for the dynamic threat landscape.

\end{abstract}

\begin{IEEEkeywords}
Cyber Threat Intelligence, Large Language Model, In-Context Learning, Cybersecurity Knowledge Graph
\end{IEEEkeywords}

\section{Introduction}
\label{sec:intro}

Modern cyberattacks are becoming increasingly complex and rapidly evolving.
Many public and commercial organizations extensively record and share cyber threat intelligence (CTI) on their platforms to combat evolving threats.
According to Gartner, CTI is defined as ``evidence-based knowledge, including context, mechanisms, indicators, implications and actionable advice, about an existing or emerging threat to assets, used to inform decisions regarding the subject’s response to that threat''~\cite{McMillan2013}.
Such knowledge is crucial for organizations to monitor the rapidly evolving threat landscape, promptly detect early signs of an attack, and effectively contain the attack with proper measures. 
Given its importance, CTI has been increasingly collected and exchanged across organizations, often in the form of Indicators of Compromise (IOC)~\cite{AcingIOC}.
IOCs are forensic artifacts of an intrusion such as virus signatures, IPs/domains of botnets, MD5 hashes of attack files, etc.
However, recent studies~\cite{AcingIOC, tounsi2019cyber} showed that knowledge offered by IOCs is rather limited, which covers only a limited set of knowledge and has a short lifespan.

Recognizing the limitations of IOCs, recent research has shifted towards automatically extracting richer knowledge from textual threat descriptions in CTI reports (i.e., CTI text). These reports, such as security blog articles~\cite{threatpost, darkreading} and news~\cite{thehackernews, securityweek}, are produced by security researchers and practitioners and published on websites, summarizing threat behaviors in natural language. Besides IOCs, these reports contain various other cybersecurity entities, such as malware, vulnerabilities, and attack techniques, as well as their interactions and dependencies. This knowledge is crucial for building a comprehensive cyber threat profile.

Several approaches have been proposed for automatically extracting security knowledge from CTI and constructing a \emph{cybersecurity knowledge graph (CSKG)}. 
Syntax-parsing-based approaches~\cite{ttpdrill,threatraptor,AcingIOC} leverage fixed dependency rules and hand-crafted dictionaries to parse the grammatical structure of sentences and extract key subject-verb-object triplets. 
Fine-tuning-based approaches~\cite{extractor,AttacKG,LADDER} leverage pre-trained transformers and fine-tune them on labeled CTI datasets to identify semantic roles and extract entities and relations. 
However, these methods suffer from several \emph{key drawbacks}, particularly facing the evolving threat landscape: 

(1) \textbf{\emph{Lack of flexibility and generalizability:}} Many of these methods are tailored to specific cybersecurity ontologies, focusing on a fixed set of entities and relation types. They are difficult to generalize to new ontologies and emerging threats and terminologies. Fixed rules have limited flexibility to adapt to new patterns and require manual creation and maintenance. Model fine-tuning, however, requires a large amount of labeled CTI corpus data. Such data is scarce in security, especially for new threats that lack annotations.

(2) \textbf{\emph{Information inaccuracy and incompleteness:}} Due to the peculiarities of the security context and the lack of deep analysis, these methods often generate low-quality CSKGs that are incomplete, inaccurate, and disconnected. \cref{fig:mot} shows example CSKGs generated by three representative methods for a real-world CTI report. We can observe issues including incomplete entities, misidentified entity boundaries, misaligned entities, missing links, etc. These low-quality CSKGs limit the ability to obtain a comprehensive threat profile, hindering the effective use of CTI to enhance defensive measures.

These limitations highlight the need for a \emph{paradigm shift} in CTI knowledge extraction that enables accurate knowledge capture in data-limited environments while adapting to evolving threats. Recent advancements in LLMs have demonstrated strong capabilities in various natural language tasks~\cite{chang2024survey}, shifting the focus from fine-tuning to in-context learning (ICL), which requires minimal annotated data and no parameter updates. However, ICL strategies vary in performance, from state-of-the-art to suboptimal~\cite{GoodICLExamples}. To address this, we conducted thorough experiments to identify optimal ICL settings for CSKG construction. With the optimized ICL strategy, LLMs can effectively learn from a few examples and adapt to new tasks with stability and high performance without requiring model weight updates.

\paragraph{Contributions} We present \tool, an LLM-powered framework for automated CTI knowledge extraction and CSKG construction from CTI reports. Unlike existing methods limited by generalizability and data demands, \tool introduces an \emph{optimized-ICL-based pipeline} for data-efficient inference, enabling precise extraction of diverse cybersecurity entities and relations while \emph{adapting to various ontologies}.
In addition, \tool refines the extracted knowledge to enhance the canonicalization and completeness of the resulting knowledge graph.
As shown in \cref{fig:mot}, the CSKG constructed by \tool has significantly higher quality compared to existing approaches.

\tool leverages the ICL paradigm of LLMs to extract entity-relation triplets (i.e., $\langle$head entity, relation, tail entity$\rangle$) by analogizing similar demonstration examples in the prompt construction, eliminating the need for large amounts of training data or extensive model tuning.
Unlike multi-round dialogue approaches, \tool performs end-to-end extraction of triplets in a single step, significantly reducing inference token costs.
To ensure the high quality of the extracted knowledge, \tool employs a carefully designed prompt template and an optimal demonstration retrieval strategy for automatic prompt construction.
This prompt construction also incorporates the defined ontology for the task domain.
Different ontologies can be easily swapped in, and with just a few demonstration examples, \tool can automatically bootstrap and adapt to new threats and tasks.

To canonicalize the knowledge and remove redundancy in entities, we designed a \emph{hierarchical entity alignment} technique, which consists of two phases.
In coarse-grained entity grouping, \tool assigns entity types to each entity in the extracted triplets using LLM’s ICL and groups entities within the same type. This ensures preliminary categorization and prevents the merging of textually similar entities that belong to different types.
In fine-grained entity merging, \tool calculates the semantic similarity among the grouped entities and merges those with high similarity.
With this hierarchical approach, \tool avoids the high costs of querying LLMs for each entity pair's similarity.

To further complete the CSKG with implicit relations for distant entities, we designed a \emph{long-distance relation prediction} technique.
First, entities with the highest degree centrality in a subgraph are selected as the central nodes of that subgraph.
Then, \tool leverages ICL to predict implicit relations 
among these central nodes 
to infer connections among the disjoint subgraphs.

\paragraph{Evaluation}
We conducted comprehensive evaluations using 150 CTI reports from 10 well-recognized CTI sharing platforms~\cite{avertium, bleepingcomputer, darkreading, googletag, securityweek, symantec, thehackernews, threatpost, trendmicro, unit42}. \tool achieved F1-scores of 87.65\% in \prefix triplet extraction, 89.94\% in coarse-grained entity grouping, 99.80\% in fine-grained entity merging, and 90.99\% in relation prediction. Qualitative analysis showed that \tool constructs more comprehensive and interconnected CSKGs compared to TTPDrill~\cite{ttpdrill}, EXTRACTOR~\cite{extractor}, and LADDER~\cite{LADDER}. Quantitatively, \tool outperforms EXTRACTOR by 25.36\% in F1-score for \prefix triplet extraction and LADDER by 19\% in \prefix entity extraction. We also explored various prompting strategies and four backbone models (closed-source models: GPT-3.5 and GPT-4; open-source models: Llama3 and QWen2.5) to identify the optimal ICL paradigm for CTI knowledge extraction, providing valuable insights for future research.
\tool's code and data are available at \href{https://ctinexus.github.io/}{https://ctinexus.github.io/}.
\section{Background and Motivating Example}
\label{sec:bg}

\subsection{Cyber Threat Intelligence}
\label{subsec:cti}

Although crowd-sourced CTI reports provide valuable information, their unstructured format significantly hinders their effectiveness. 
As the number and complexity of cyberattacks increase, the textual CTI descriptions have also expanded, creating an urgent need for automated information extraction from CTI~\cite{rahman2023attackers}.
The extracted knowledge can be used to construct cybersecurity knowledge graphs (CSKGs), where nodes represent entities and edges represent relations. 
Compared to unstructured CTI text, CSKGs provide a holistic profile for cyber threats, offer better visualization, and are more amenable to integration into downstream applications.
The construction of a CSKG typically follows an ontology, which specifies the entity types and their allowed relations.
Despite the development of various security ontologies~\cite{STUCCO,Malont,UCO} covering different aspects of threats, the rapid evolution of threats makes it nearly impossible to maintain a universal, comprehensive ontology.
This underscores the need for CTI knowledge extraction approaches that can adapt to different ontologies and emerging threats with minimal transition effort.

\begin{figure*}[t]
    \centering
    \includegraphics[width=\textwidth]{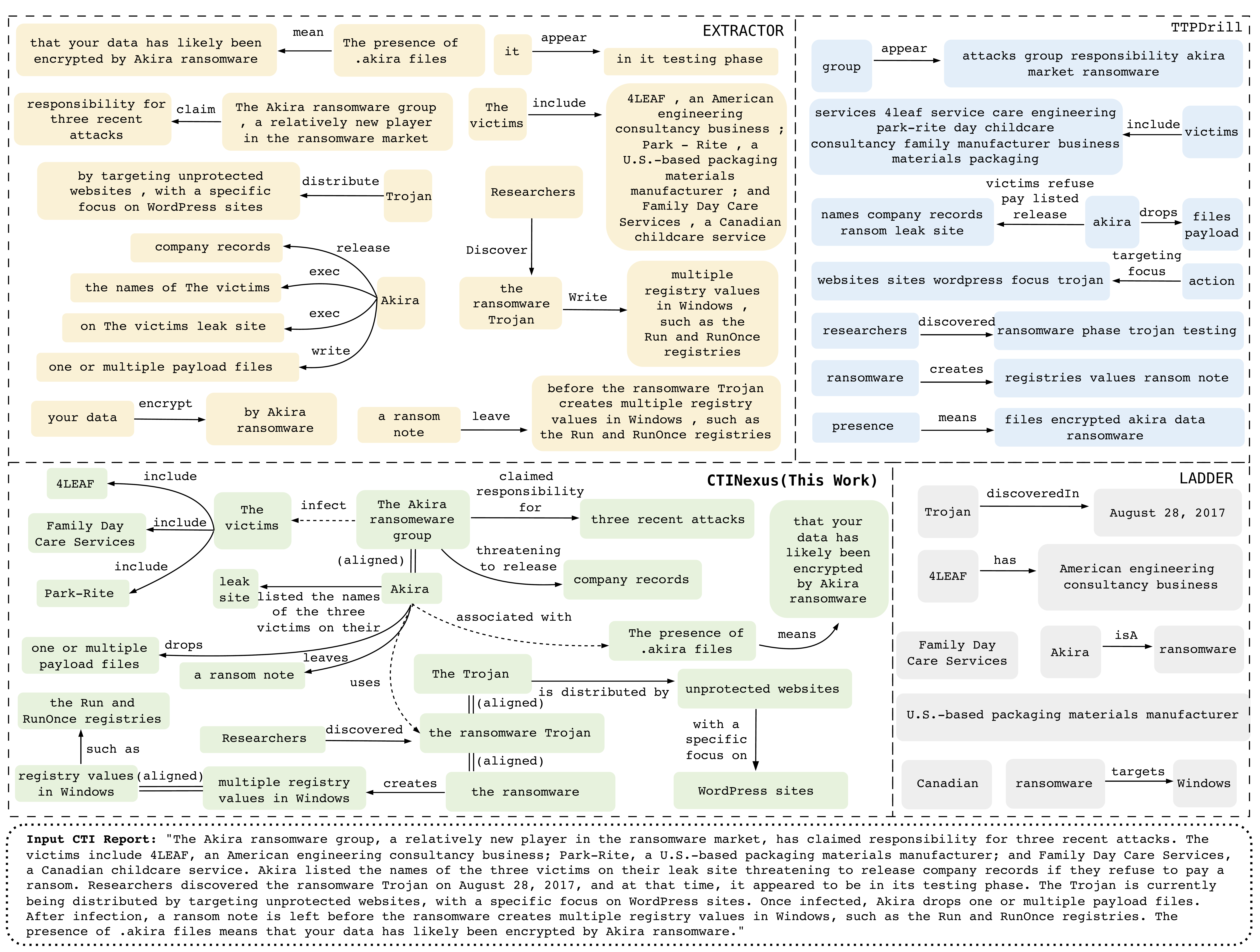}
    \caption{CSKGs extracted by EXTRACTOR, TTPDrill, LADDER, and \tool for a real-world CTI report. EXTRACTOR,TTPDrill, and LADDER tend to produce incomplete and fragmented subgraphs, lacking comprehensive contextual connections.
    In contrast, \tool constructs a more integrated and comprehensive CSKG, with key information extracted and entities linked, providing a clearer and more complete representation of the threat profile.}
    \label{fig:mot}
    \vspace{-1ex}
\end{figure*}

\subsection{Limitations of Existing Approaches}
\label{subsec:limitations}

Existing CTI knowledge extraction approaches face several fundamental challenges in adapting to the rapidly evolving threat landscape.
Existing approaches follow two paradigms: syntax parsing-based and fine-tuning-based. 
\emph{Syntax parsing-based methods} leverage typed dependency rules to analyze the grammatical structure of a sentence and extract subject-verb-object (SVO) triplets.
For example, TTPDrill~\cite{ttpdrill} extracts subject entities and verb relations in CTI-related sentences as threat actions.
iACE~\cite{AcingIOC} extracts verb relations between IOCs and context terms.
ThreatRaptor~\cite{threatraptor} extracts verb relations between subject IOC and object IOC.
However, syntax parsing-based methods have \emph{two main drawbacks:}

\begin{itemize}[leftmargin=*,itemsep=6pt, topsep=6pt]
    \item \emph{Domain complexity:} The grammatical rules can apply to any domain. However, CTI text has several peculiarities that can confuse syntax parsing, leading to inaccurate extraction. Cybersecurity entities can contain special characters, such as dots in IPv4 addresses, underscores in file names, and slashes in file paths. These special characters can confuse basic NLP modules, like sentence segmentation and tokenization, which syntax parsing relies upon. 

    \item \emph{Static nature:} These methods rely on fixed syntax rules and predefined dictionaries to filter out irrelevant information and canonicalize extracted information. For example, TTPDrill maps extracted SVOs to a curated list of threat action terms, while ThreatRaptor uses a dictionary to canonicalize the extracted relation verbs. Keeping up with the evolving threat landscape requires continuous updates and maintenance of these rules and dictionaries, which is hard to scale.
\end{itemize}

On the other hand, \emph{fine-tuning-based approaches} fine-tune pre-trained neural networks on annotated CTI domain datasets to perform named entity recognition (NER) and relation extraction (RE).
For example, EXTRACTOR~\cite{extractor} fine-tunes a pre-trained BERT~\cite{shi2019simple} model with thousands of annotated CTI sentences, to perform semantic role labeling to extract subjects, objects, and verb actions.
AttacKG~\cite{AttacKG} fine-tunes a pre-trained model in the SpaCy library~\cite{spaCy} to recognize entities and extract dependencies.
LADDER~\cite{LADDER} fine-tunes different pre-trained transformers, including BERT, RoBERTa, and XML-RoBERTa, on their custom datasets annotated according to their own ontology for performing NER and RE.
ThreatKG~\cite{threatkg-lamps} trains domain-specific BiLSTM and PCNN-ATT models for extracting security entities and relations. 
However, fine-tuning-based methods also have \emph{several drawbacks:}

\begin{itemize}[leftmargin=*,itemsep=6pt, topsep=6pt]
    \item \emph{Resource requirement}: Model training and fine-tuning require large amounts of labeled data (i.e., annotated CTI text corpora), and the labeling needs to be aligned with the targeted ontology. Such annotations are expensive to obtain, especially for emerging threats.
    Additionally, fine-tuning can be computationally expensive if the backbone model contains lots of parameters.    

    \item \emph{Ontology lock-in}: Since the models are fine-tuned on datasets annotated using a specific ontology, they are \emph{\textbf{difficult to generalize to new ontologies}} that cover different entities and relations. Transferring to other ontologies would require reannotating vast data and retraining the models, which is very costly.    
\end{itemize}

\subsubsection{Motivating Example}

We further investigate the quality of the constructed CSKG by existing approaches using a real-world CTI report. \cref{fig:mot} illustrates a snippet of the report titled ``RANSOMWARE - AKIRA AND RAPTURE'' published on May 9, 2023, by Avertium~\cite{avertium}. 
The report provides rich information about the new Akira ransomware group. 
We run this CTI text snippet with three representative approaches, TTPDrill, EXTRACTOR, and LADDER using their released implementations~\cite{TTPDrill-code, EXTRACTOR-code, LADDER-code}. 
\cref{fig:mot} shows their constructed CSKGs.
\textbf{\emph{We observe that the quality of CSKGs is very low.}}

\begin{itemize}[leftmargin=*,itemsep=6pt, topsep=6pt]
    \item \emph{Some triplets have wrong directions.} For example, in EXTRACTOR, `` ransom note'' is extracted as the subject of ``leave'', whereas it should be the object. 

    \item \emph{Many extracted entities have poor quality.} Some are not meaningful, such as ``presence'' extracted by TTPDrill. Others include unnecessary words or combine multiple distinct entities; for example, TTPDrill incorrectly extracts ``registry values'' and ``ransom note'' together when they should be separate. Similarly, in EXTRACTOR, the victim entities are not properly distinguished and should be individually separated. Although LADDER's extracted content is of higher quality compared to TTPDrill and EXTRACTOR, it often lacks completeness. For instance, in the context where a ``Trojan'' targets ``WordPress sites'', LADDER only extracts ``WordPress'' thereby omitting contextual information from the original phrase.

    \item \emph{Entities are not aligned}. For example, in EXTRACTOR, ``Trojan'' and ``the ransomware Trojan'' refer to the same object and should be merged or associated. The same issue is observed in TTPDrill and LADDER.

    \item \emph{Some critical relations are missing}. In the text, ``the Akira ransomware group'' uses the ``ransomware Trojan'' to launch the attack. However, since these two entities are mentioned in different sentences without explicit relational indicators, all approaches fail to infer the relationship between them.
    
\end{itemize}

As shown in \cref{fig:mot}, the CSKG constructed by \tool is comprehensive, well-connected, and has much better quality, addressing all previous drawbacks. 
By leveraging the in-context learning of LLMs, the construction of such a CSKG does not rely on large amounts of training data and can adapt to different ontologies.
We describe our approach in \cref{sec:design}.

\begin{figure*}[t]
    \centering
    \includegraphics[width=0.9\textwidth]{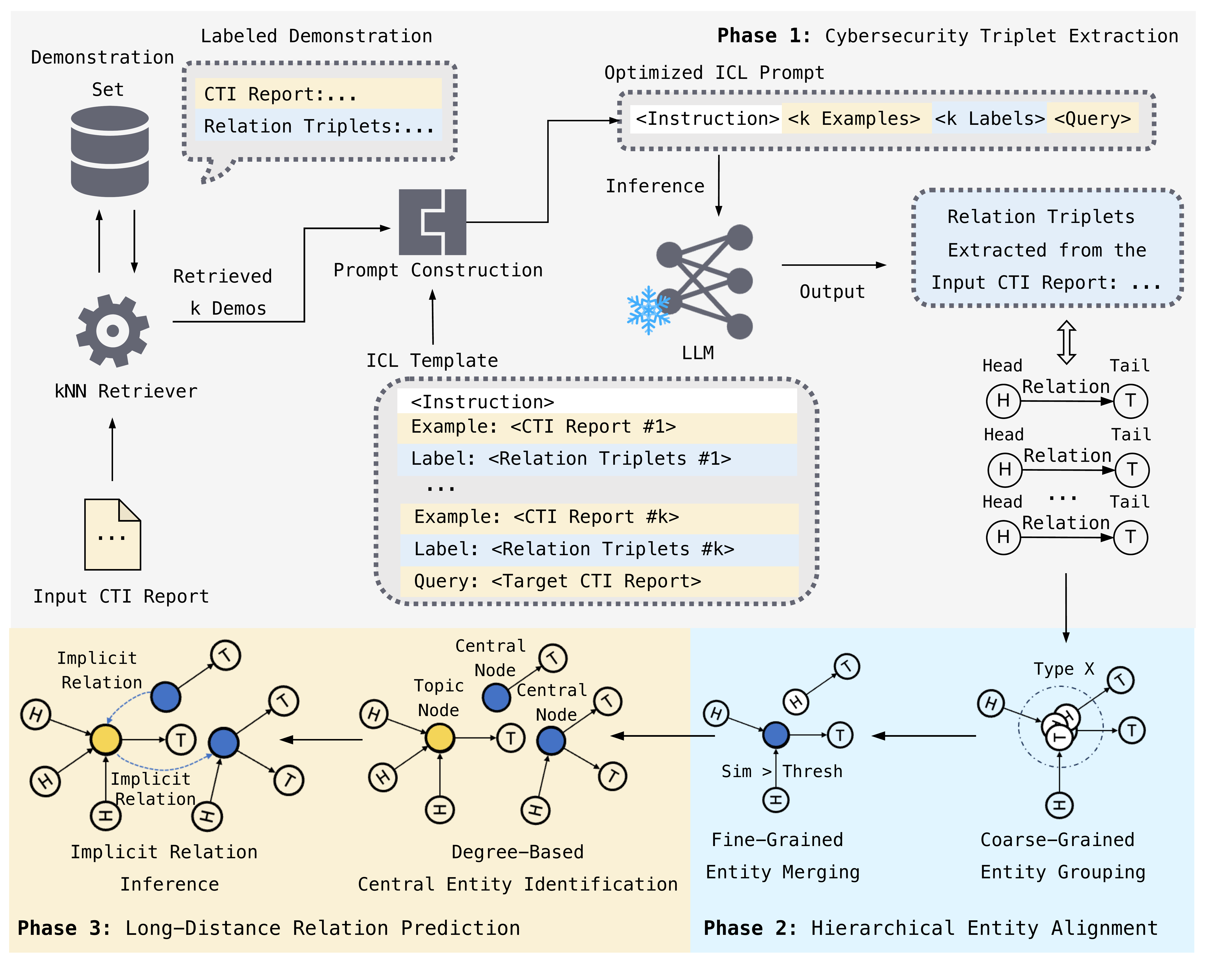}
    \caption{Overview of \tool. \tool comprises three phases. Phase 1, \emph{Cybersecurity Triplet Extraction}, enables end-to-end extraction of \prefix triplets using in-context learning of LLM. Phase 2, \emph{Hierarchical Entity Alignment}, reduces the redundancy of CSKG through coarse-grained grouping and fine-grained clustering. Phase 3, \emph{Long-Distance Relation Prediction}, connects disjoint subgraphs by identifying central nodes and performing relation inference.
    }
    \label{fig:Overview}
    \vspace{0ex}
\end{figure*}

\subsection{Large Language Models (LLMs)}

Recently, LLMs have shown emergent abilities to learn from just \emph{a few demonstration examples} in the prompt, a paradigm known as in-context learning (ICL)~\cite{Dong2022}. In the ICL paradigm, the prompt input to the LLM typically includes three components: (1) an instruction specifying the task, (2) several demonstration examples containing ground truth to provide task-specific knowledge, and (3) a query to the LLM with the expectation of an appropriate answer. This allows LLMs to adapt to new tasks with minimal cost using task-specific prompts and demonstration examples.
Multiple studies have shown that LLMs perform well in various tasks under ICL, such as fact retrieval~\cite{wang2023survey} and mathematical reasoning~\cite{imani2023mathprompter, ahn2024large}. Additionally, LLMs have shown promise in different cybersecurity tasks, such as vulnerability detection~\cite{fang2024llm, lu2024grace}, patch generation~\cite{kulsum2024case}, and software fuzzing~\cite{xia2024fuzz4all, meng2024large}. 
However, the use of LLMs for CTI knowledge extraction and CSKG construction remains largely underexplored.

\section{Overview} 

\cref{fig:Overview} illustrates \tool.
\tool introduces an ICL-based approach for data-efficient CTI knowledge extraction and CSKG construction.
Unlike previous methods, \tool eliminates the need for extensive data annotations and parameter tuning, facilitating generalization to various ontologies. 
\tool focuses on constructing connected and comprehensive CSKGs, enabling entity alignment and long-distance relation inference. 
\tool includes three phases.

\emph{Phase 1:} Given a CTI report, \tool first extracts entity-relation triplets that align with the task ontology. 
The kNN-based demonstration retriever embeds the report and the candidate reports in the demonstration set into a high-dimensional latent space.
The retriever then selects the top-$k$ candidates with the highest similarity scores.
The selected demonstrations are fed into an automatic prompt construction module to create a customized prompt for the current report. As illustrated in \cref{fig:Overview}, our prompt template consists of three sections: an instruction describing the task, a query containing the input CTI report, and demonstration examples arranged in a specific order.
\cref{fig:ICL-prompt} illustrates our carefully designed instruction. 
\textbf{\emph{Note that the ontology is incorporated into the instruction}}. This design allows different ontologies to be easily switched, and our automatic prompt construction module will create a prompt specifically for this ontology and report, enhancing knowledge extraction performance.

\emph{Phase 2:} With the extracted triplets, \tool removes redundancy by merging entities that refer to the same cybersecurity object using a hierarchical approach.
The coarse-grained entity grouping module assigns types to entities using an automatically populated ICL prompt template, as illustrated in \cref{fig:EA}. The instruction incorporates the ontology that defines possible entity types. The demonstration examples show how to label each entity in the triplet. The query includes all the triplets to be typed. 
Entities assigned the same type are grouped together.
Next, the fine-grained entity merging module embeds all entities within each group and merges those that exceed a predefined similarity threshold into a single entity.

\emph{Phase 3:}
To infer missing links between distant entities, \tool performs long-distance relation prediction. The central entity identification module selects a central node in each connected subgraph based on the node’s degree centrality. Among central nodes, the module then selects a topic node with the highest importance, which serves as the main subject of the report. The central nodes and the topic node are passed to the ICL-enhanced relation prediction module to infer their implicit relationships.
\tool automatically constructs an ICL prompt (illustrated in \cref{fig:LP}) to perform this inference.

\section{Design of \tool}
\label{sec:design}

\subsection{CSKG Ontology}
\label{subsec:problem}
\begin{figure*}[t]
    \centering
    \includegraphics[width=\textwidth]{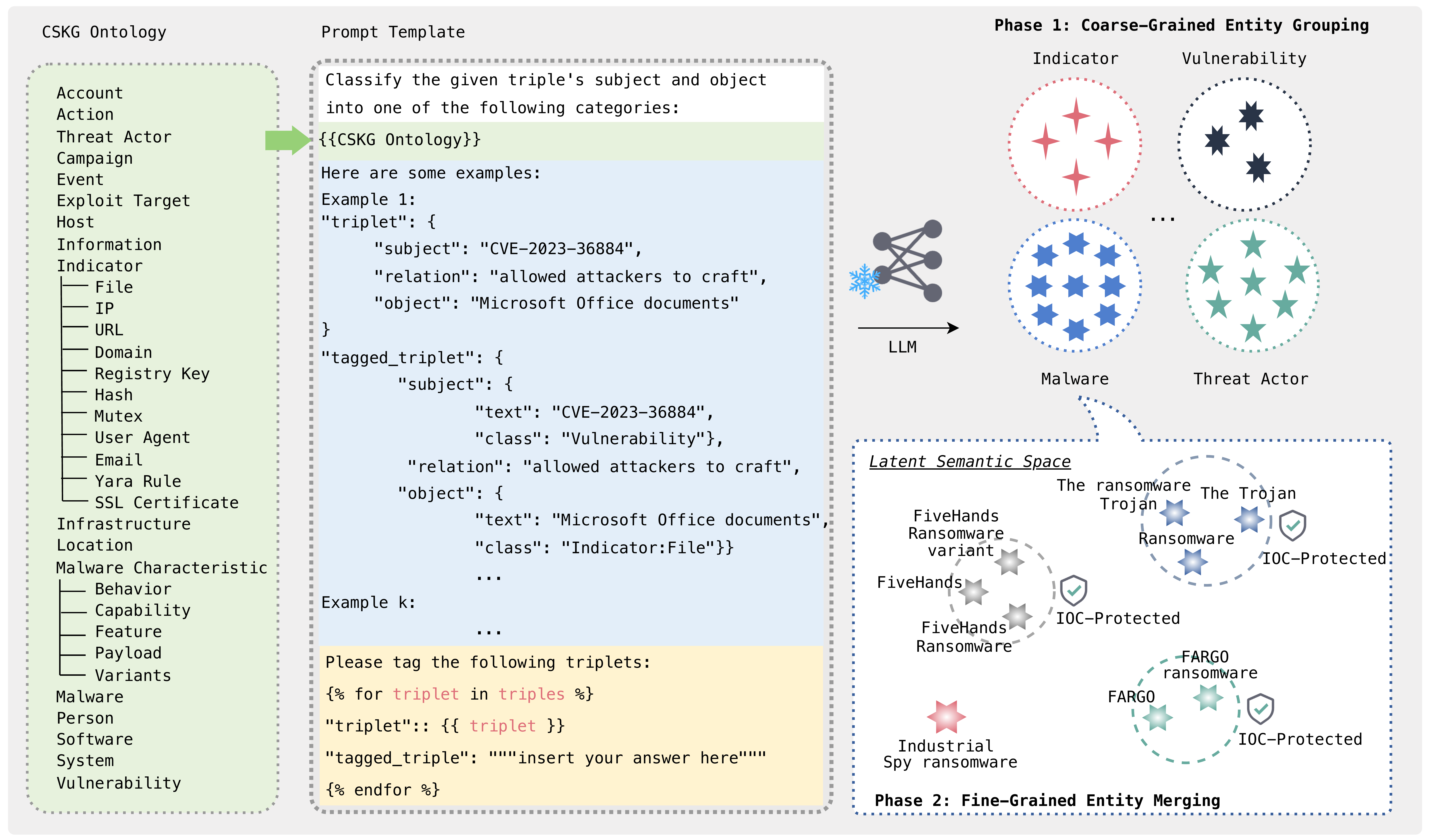}
    \caption{The design of \tool's hierarchical entity alignment. The coarse-grained entity grouping phase populates an ICL prompt to assign entity types to the extracted triplets according to the applied ontology. Entities with the same type are grouped together. The fine-grained entity merging phase then uses an embedding-based technique to merge semantically similar entities within each group based on a predefined similarity threshold. During this phase, IOC protection is enforced to prevent erroneously merging semantically similar but conceptually distinct IOC entities.
    }
    \label{fig:EA}
    \vspace{-0ex}
\end{figure*}

We choose MALOnt for the current implementation, as MALOnt~\cite{Malont} is the most comprehensive among open-source ontologies, featuring 33 entity types (17 types and 16 sub-types) and 27 relation types. 
MALOnt covers a broad range of entities, such as Account, Action, Threat Actor, Campaign, Event, Exploit Target, Host, Information, Infrastructure, Location, Malware, Person, Software, System, and Vulnerability, with detailed sub-types under Indicator and Malware Characteristics. 
However, note that \toolno’s ICL-based pipeline eliminates the need for parameter tuning on large, ontology-specific training sets, largely simplifying generalization to other ontologies. 
If downstream applications require ontologies not covered by MALOnt, \tool can easily switch to a different ontology. This only requires a few demonstration examples aligned with the new ontology for each ICL task, and the ontology defined in a JSON format incorporated in the prompts (illustrated in \cref{fig:ICL-prompt,fig:EA}). 
If the new ontology is a subset of MALOnt (which is already quite comprehensive), \tool can directly adapt by simply removing unrequired entity types without further actions.

\subsection{Cybersecurity Triplet Extraction}
\label{subsec:TE}

\begin{figure*}[hbt!]
    \centering
    \includegraphics[width=\textwidth]{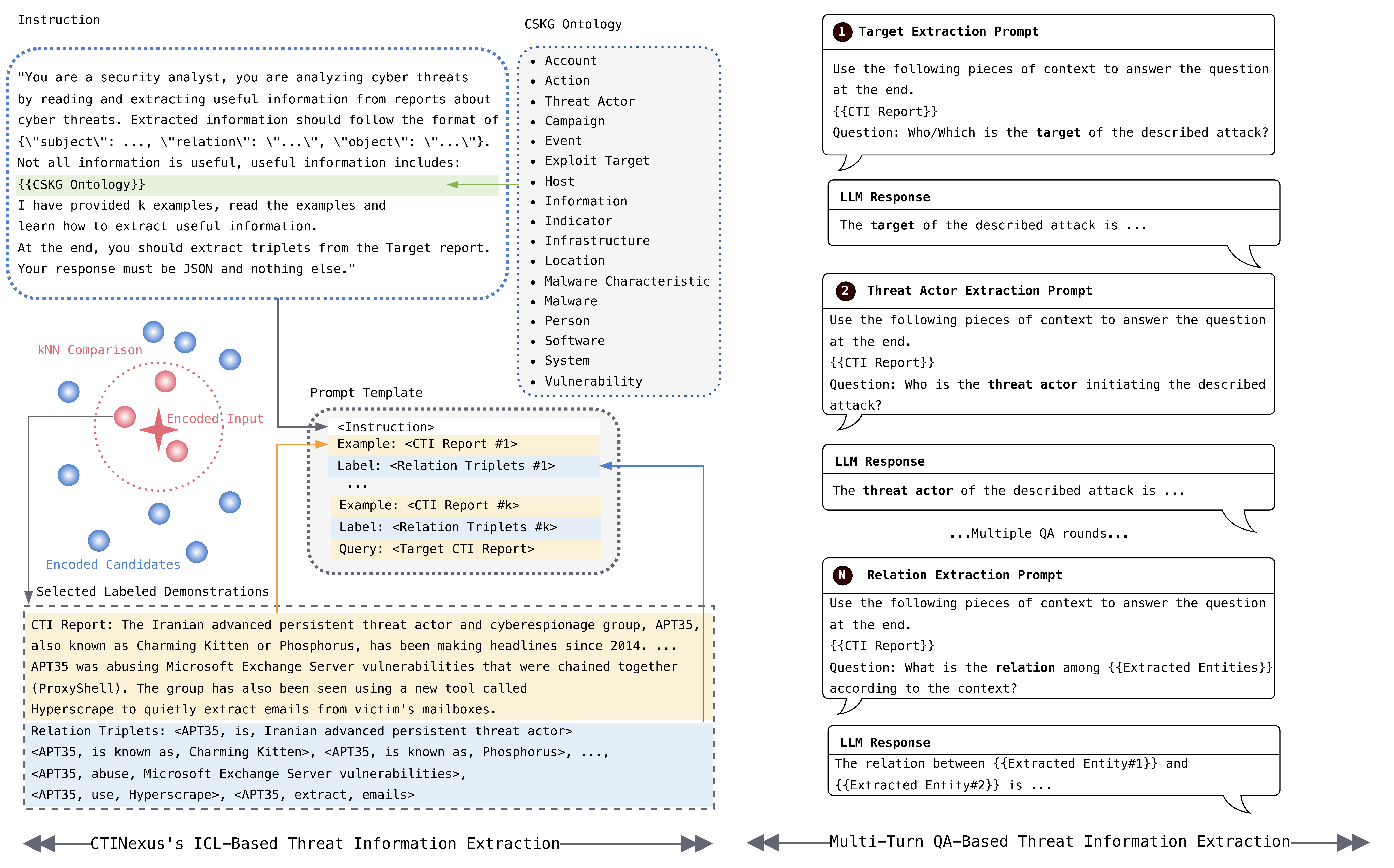}
    \caption{
    Comparison of \tool's ICL-based CTI knowledge extraction (left) and a multi-turn QA-based extraction (right).
    \tool consolidates task descriptions (including applied ontology), $k$ selected demonstrations, and query into a single instruction for efficient~\prefix triplet extraction. In contrast, the multi-turn QA paradigm requires multiple rounds of conversations with multiple prompts to extract different entities and relations, which is inefficient. 
}
    \label{fig:ICL-prompt}
    \vspace{-0ex}
\end{figure*}

Given that CTI text may contain diverse relations and we want the approach to be adaptable to emerging threats, we formulate the \prefix triplet extraction module in our pipeline as a \emph{semi-open} extraction problem: Entity types follow MALOnt, as its coverage is already comprehensive, while relation extraction is modeled as open RE to maximize the coverage.
These approaches transform information extraction tasks into multi-turn question-answering, leveraging the conversational capabilities of LLMs.
\cref{fig:ICL-prompt} illustrates this paradigm. 
This method involves creating multiple questioning prompts for each information type and refining the responses. However, applying this multi-turn QA formulation to cybersecurity entity and relation extraction requires numerous lengthy prompts due to the extensive cybersecurity ontology that could contain many entity classes. 
For $N$ entities in the input CTI, $\frac{N(N-1)}{2}$ prompts are needed to extract relations between identified entities, leading to repetitive content and significant token waste, hindering scalability. 
Additionally, the multi-turn paradigm suffers from confirmation bias~\cite{confirmationBias}, as LLMs may confirm with a non-existing relation after several rounds of dialogue. 
In~\cref{para:prompt_design}, we present our evaluation of prompt formulation and strategy that underpin \toolno's superiority over the multi-turn QA formulation and baseline prompt designs.

\paragraph{ICL prompt template}
\label{para:icl_template}
To improve efficiency and reduce confirmation bias, we develop a kNN-enhanced ICL paradigm that completes the \prefix triplet extraction process with only one LLM query. 
As illustrated in \cref{fig:ICL-prompt}, \tool extracts all \prefix triplets by automatically populating a comprehensive ICL prompt template, which comprises the following components:

\begin{enumerate}[label=(\arabic*), leftmargin=*, itemsep=6pt, topsep=6pt]

    \item \emph{Instruction}: The instruction specifies the task, the applied ontology, and the required format for the extracted triplets. Instruction design is critical in LLMs, as an unclear definition of the task can severely degrade the performance. We carefully designed several versions of the instruction and identified the one presented in \cref{fig:ICL-prompt} as the most effective.

    \item \emph{Demonstrations}: Top-$k$ most relevant examples are retrieved using the demonstration retriever. Each consists of a CTI report annotated with the security triplets. These examples are ordered in ascending similarity to the input query based on our findings described in \cref{subsub:example-perm}.

    \item \emph{Query}: The input CTI text that needs to be analyzed.
\end{enumerate}

\paragraph{kNN-based demonstration retriever}
\label{para:kNNRtr}
Multiple studies~\cite{qin2023context, GoodICLExamples} have shown that prompt examples selection can significantly affect LLM's ICL capacity. One approach for selecting demonstration examples involves training a proxy LM to score candidates in the demonstration set~\cite{xu2024context}. However, this method requires large amounts of labeled data, which conflicts with our goal of designing a data-efficient solution. Recently, a k-nearest neighbors (kNN) method for selecting the most relevant demonstration examples based on semantic similarity has proven effective~\cite{GoodICLExamples}. This method requires no dataset annotation or model tuning, making it ideal for our purposes. Specifically, we compute high-dimensional embeddings for the query and all candidate demonstrations using a pre-trained embedding model. Among the models explored, text-embedding-3-large yielded the best performance. We then calculate the cosine similarity between the query embedding and each candidate demonstration’s embedding, selecting the top-$k$ most similar candidates.

Several studies \cite{GoodICLExamples,Dong2022,min2022rethinking} have pointed out that the order of demonstration examples can also affect the performance of ICL. 
In particular, the model's prediction often exhibits a recency bias~\cite{liu2024lost}, meaning that LLMs tend to pay more attention to the demonstration placed near the query. Also, kNN similarity indicates that if the demonstration is more similar to the query, LLMs can better analogize it.
To investigate the impact of demonstration order in the CTI domain, we evaluated various permutations, including random, ascending, and descending orders (\cref{subsub:example-perm}). Our findings indicate that arranging the demonstration examples in ascending order of their similarity to the query yields the best performance.
This confirms the recency bias phenomenon in our scenario, as the demonstration example most similar to the query is placed at the bottom of the list, closest to the query.

\subsection{Hierarchical Entity Alignment}
\label{subsec:EA}

\begin{figure*}[hbt!]
    \centering
    \includegraphics[width=\textwidth]{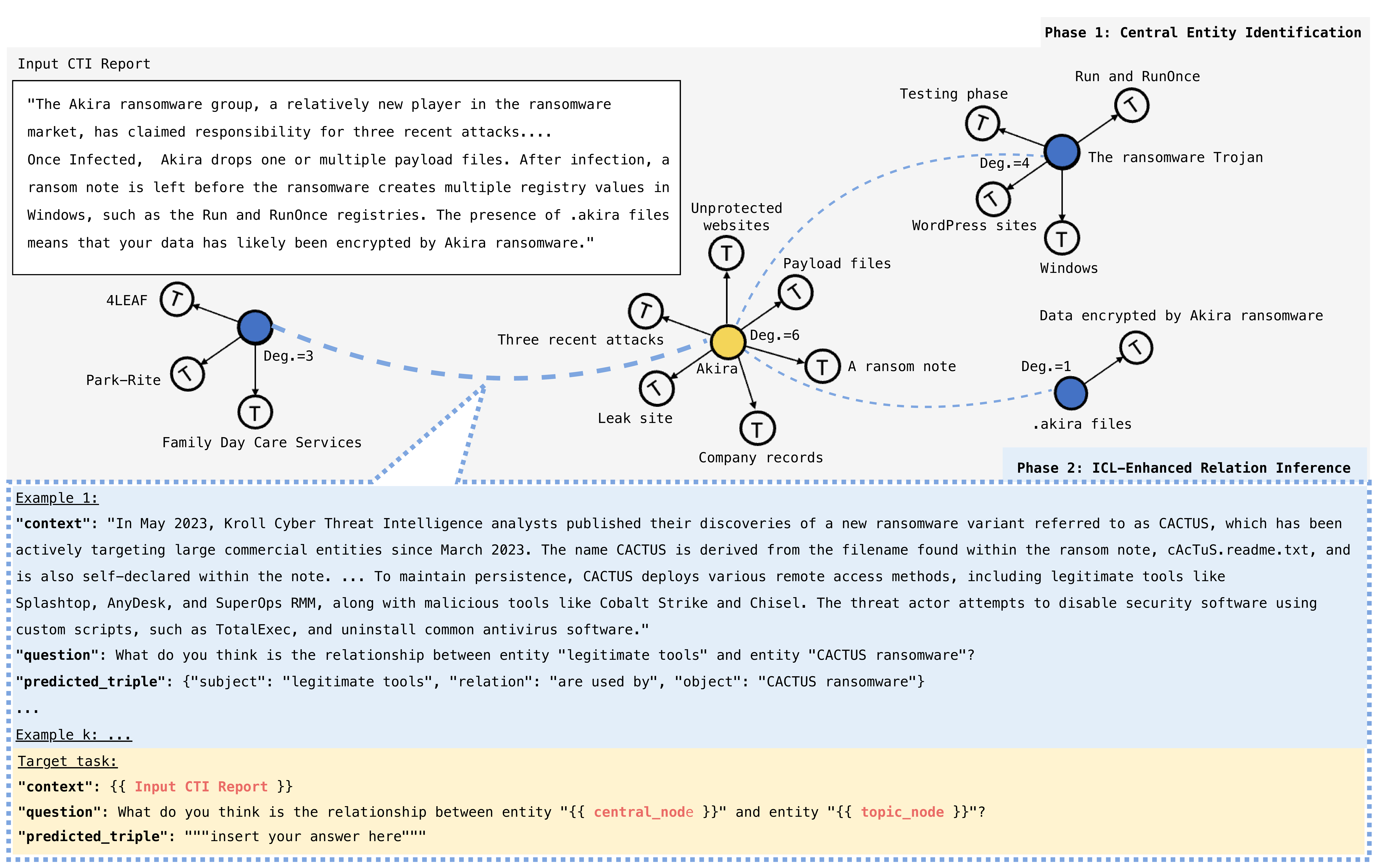}
    \caption{The design of \tool's long-distance relation prediction. Phase 1 selects central entities (blue) and the topic entity (yellow) from separate subgraphs based on their degree centrality. Phaes 2 populates an ICL prompt to infer implicit relations between each central entity and the topic entity.
    }
    \label{fig:LP}
    \vspace{-0ex}

\end{figure*}

Entity alignment identifies entities with different mentions that refer to the same real-world object, a key area in knowledge graph research~\cite{abu2021domain}. Aligning these mentions integrates sub-graphs containing complementary knowledge, enhancing the comprehensiveness of the knowledge graph.
Traditional entity alignment techniques rely on heuristics like string matching and structural similarities, which fail to capture the underlying semantics or context of entities and have limited accuracy.
Recent studies~\cite{sun2018bootstrapping,trisedya2019entity,pei2019semi} have adopted deep learning-based methods to learn vector representations (i.e., embeddings) of entities, achieving better accuracy. However, embedding-based techniques face unique challenges in our problem domain. 
In CTI text, entities with similar embeddings may refer to different concepts, e.g., ``.akira files'' (an IOC) and ``Akira'' (a threat actor).
Besides, comparing the semantic distance between every pair of entities has a computational complexity of $n^2$, where $n$ is the total number of entities. This is inefficient when $n$ becomes large.

To address these challenges, we perform entity alignment in a \emph{hierarchical} way. 
The coarse-grained entity grouping module leverages LLM's ICL ability to assign types to entities. Entities assigned the same type are then grouped together as potential candidates for alignment, narrowing the scope for later fine-grained merging.
\cref{fig:EA} illustrates our prompt template.
\tool automatically creates a customized prompt by assembling $k$ carefully annotated demonstration examples. 
Each demonstration example contains an untagged triplet and a tagged triplet with subject and object entities assigned type labels. The query part automatically traverses all triplets generated by the triplet extraction phase. For each triplet, we add a placeholder, ``tagged\_triplet'': ``insert your answer here'' to follow the format provided in the demonstration examples, better guiding the LLM to correctly fill in the answers.

For entities within each group, the fine-grained entity merging module uses an embedding-based technique to merge entities with similar semantic representations. The embedding model is central to this procedure, as its generated embeddings are used to determine the semantic closeness of entities. We evaluated state-of-the-art, general-purpose text embedding models of various sizes (i.e., text-embedding-3-small, text-embedding-3-large) for this task. Since these models are not specifically pre-trained on a cybersecurity corpus, we also experimented with a security-specific embedding model, SecureBERT~\cite{securebert}, which has been pre-trained on millions of cybersecurity websites, articles, and books.
Another aspect to consider is the similarity threshold (degree of closeness) for determining alignment. To find the optimal threshold value, we experimented with common threshold values in semantic similarity comparison~\cite{pilehvar2013align}: 0.4, 0.5, 0.6, and 0.7. Our results indicated that 0.6 is the most effective value.
Detailed results and discussions are in \cref{eval:entity-alignment}.

To further ensure safe merging, we introduce an IOC protection mechanism that prevents merging semantically similar but conceptually distinct IOCs sharing the same general type (e.g., CVE-2023-23397 vs. CVE-2023-23392). Specifically, we use a curated set of regular expressions to identify IOC patterns (e.g., CVE strings, IP addresses, file hash, etc.) and isolate them from being merged with other mentions.

\subsection{Long-Distance Relation Prediction}

After entity alignment, the triplets form a set of disconnected subgraphs, leaving implicit relations between distant entities unidentified. Previous methods primarily rely on graph structure learning and graph neural networks~\cite{zhu2021survey, jin2020graph} to perform link prediction. However, these methods require large amounts of annotated graph data for model training. Additionally, in the CTI analysis domain, establishing relationships between distant cybersecurity entities requires a deep natural language understanding of their corresponding context. To make the procedure more data-efficient, we develop a long-distance relation prediction technique leveraging the ICL ability of LLMs.
\cref{fig:LP} illustrates our design.

Creating links for every pair of distant entities would introduce excessive connections, complicating the CSKG and consuming significant computational resources. 
Thus, \tool first runs a depth-first search to find all connected subgraphs.
Then, \tool leverages graph structure information to identify a central entity for each subgraph. A central entity represents the most important entity in the subgraph and will be the head for inter-subgraph connections.
In our design, we identify central entities based on their degree centrality~\cite{zhang2017degree}, which is the most widely used measure of a node's importance in a graph.
It is easy to calculate, by counting the total number of edges that a node has to other nodes.
The intuition is that an entity with the most explicit relations with other entities is more likely to be the core subject of that part of CTI text.
Among all identified central entities, we further identify a topic entity, which is the one with the highest degree, representing the core subject of the entire CTI report.
Specifically, we consider both incoming and outgoing edges when calculating degree centrality to identify the central identity. If multiple entities have the same highest score, we further prioritize out-degree over in-degree, as subjects in triplets (e.g., ``Androxgh0st'' in <``Androxgh0st'', ``targets'', ``.env files''>) are generally more important than objects. 
If there is still a tie, they are all determined as central entities.
We follow the same procedure for identifying the topic entity. 
In the example shown in \cref{fig:LP}, there are five subgraphs. We identify the following central entities: ``Victim'', ``Akira'', ``the ransomware Trojan'', ``Akira ransomware group'', and ``.akira files''.
We select ``Akira'' as the topic entity, which has the highest degree centrality score of $6$.
These central entities and the topic entity are then fed into the next module for relation inference.

The ICL-enhanced relation prediction module leverages ICL of LLM to infer implicit relations between each central entity and the topic entity, creating inter-subgraph connections. \cref{fig:LP} illustrates our prompt template. For each central entity, \tool automatically creates a customized prompt by assembling $k$ fixed, carefully annotated demonstration examples, similar to the entity alignment process.
The prompt template consists of two sections: a demonstration section (blue) and a query section for the target task (yellow). 
Both sections include ``context'', ``question'', and ``predicted\_triple'' components.
The ``context'' component presents the CTI report, while the ``question'' component asks the LLM about the relation between the queried central entity and the topic entity. 
The ``predicted\_triple'' component contains the annotated relations for the demonstration examples and a placeholder, ``insert your answer here'', for the queried task.
This consistent design across the three components in both the query and demonstration sections helps the LLM effectively analogize the demonstration examples, facilitating better relation inference.
\section{Evaluation}
\label{sec:eval}

To comprehensively study the performance of \tool
in various phases of CSKG construction, we set the following research questions:

\begin{enumerate}[label={\textbf{RQ\arabic*:}}, leftmargin=*,itemsep=6pt, topsep=6pt]

\item How does \tool compare with existing methods for CTI knowledge extraction?

\item How do different settings in \tool affect the cybersecurity triplet extraction?

\item How do different settings in \tool affect the entity alignment and relation prediction?

\item How well does \tool perform in end-to-end CSKG construction?

\item How well does \tool adapt to different CSKG ontologies?

\item What is the efficiency of \tool?
\end{enumerate}

\subsection{Dataset and Metrics}
\label{subsec:Dataset}
Existing datasets primarily benchmark cybersecurity triplet extraction but do not comprehensively support other tasks in our pipeline. Additionally, their CTI reports are often outdated. For instance, LADDER’s dataset~\cite{LADDER} includes reports only from 2010 to 2021. To address these limitations, we curated a new dataset specifically for evaluating \tool across cybersecurity triplet extraction, hierarchical entity alignment, and long-distance relation prediction tasks. Our dataset consists of 150 recent CTI reports (May 2023 onwards) from 10 reputable sources, such as Trend Micro, Symantec, and The Hacker News. 

\paragraph{Annotations} Annotations were performed via a structured four-step approach: (1) Annotating cybersecurity entities and their types.
(2) Identifying explicit semantic relations among entities, organized into JSON-formatted triplets. (3) Grouping entities by type and merged mentions referring to identical concepts. (4) Identifying central entities and summarizing implicit inter-entity relations.

To ensure quality and reduce annotation bias, three PhD students with expertise in threat intelligence independently conducted annotations, with the third serving as an arbiter to resolve conflicts. Inter-annotator agreement, measured by Cohen’s kappa~\cite{mchugh2012interrater}, yielded scores of 0.80 (triplet extraction), 0.78 (entity alignment), and 0.61 (relation prediction), averaging 0.73, indicating substantial agreement.
This rigorous annotation procedure resulted in 3,682 entity mentions, 2,039 unique entities, and 1,982 relationships, enabling comprehensive evaluation of \toolno’s performance in constructing accurate cybersecurity knowledge graphs

\paragraph{Metrics}\label{para:metrics}
For Phase 1 (triplet extraction) and Phase 3 (relation prediction), performance is measured at the triplet level. A predicted <subject, relation, object> counts as a true positive when the three elements semantically match a gold‑standard triplet, treating active‑ and passive‑voice variants as equivalent. Predicted triplets with no gold counterpart are false positives, and gold triplets that are not predicted are false negatives.
Phase 2 (entity alignment) is assessed at the entity level: a mention is correct if it is (i) assigned the right MALOnt type during coarse‑grained grouping, and (ii) merged with the gold mention that refers to the same cybersecurity entity during fine‑grained clustering.

\subsection{RQ1: How does \tool compare against existing CTI knowledge extraction methods?}
\label{subsec:rq1}

We evaluated \tool against two state-of-the-art baselines, EXTRACTOR~\cite{extractor} and LADDER~\cite{LADDER}, representing syntactic-analysis-based and fine-tuning-based approaches, respectively.

Several methodological challenges were addressed to enable fair comparison. For EXTRACTOR, we adapted its output to our broader ontology using \toolno's coarse-grained entity grouping module. For LADDER, we addressed two key differences: 
(1) LADDER uses a word-level annotation format, where each token is labeled with its target class. In contrast, our dataset follows an end-to-end report-to-triplet format, where the entire report is input, and the label is a set of extracted triplets.
(2) LADDER uses a simplified ontology derived from MALONT, which includes only 10 entity types, a subset of the entity types used in our ontology.
To facilitate comparison, we developed scripts to tokenize our data and convert our manually annotated datasets into LADDER's word-level format. 
To ensure a fair comparison with LADDER, we merged our training set with LADDER's in a 5:1 ratio, maintaining their original training/validation split. 
We also replaced LADDER's test set with ours to ensure consistent evaluation on the same data. 
We compare with LADDER solely on named entity extraction performance, as our method focuses on open relation extraction, while LADDER targets relation classification within fixed categories.

\cref{tab:EXTRACTOR} demonstrates that \tool outperforms EXTRACTOR in all metrics 
in \prefix triplet extraction.
The evaluation results in \cref{subsec:rq2} show that GPT-4 outperforms all other backbone models. Thus, we use GPT-4 as the default backbone model for \tool's implementation. 

This superior performance can be attributed to several factors. First, \tool leverages the robust context understanding and instruction-following capabilities of LLMs and enhances specificity with kNN-selected demonstration examples for extracting triplets. 
In contrast, EXTRACTOR employs general fine-tuning to extract semantic roles not specific to any ontology, reducing its accuracy in triplet extraction. 
Also, the CTI context introduces peculiarities that lead to errors in EXTRACTOR's semantic role labeling module, which relies on a simple BERT model. 
For instance, EXTRACTOR might extract a triplet like $\langle$``Androxgh0st malware'', ``support'', ``numerous functions capable of abusing the Simple Mail Transfer Protocol (SMTP), such as scanning and exploiting exposed credentials and application programming interfaces (APIs), and web shell deployment''$\rangle$, where the object is a long sentence not suitable as a single entity.
The object contains multiple entities due to misidentified boundaries. 
Conversely, \tool captures implicit meanings and transforms phrases to be more suitable as entities, resulting in a triplet like $\langle$``Androxgh0st malware'', ``supports'', ``functions abusing SMTP''$\rangle$.

\cref{tab:LADDER} demonstrates that \tool outperforms LADDER in F1-Score, precision, and recall by 26.7\%, 17.5\%, and 19.5\%, respectively. 
Specifically, LADDER achieved an F1-Score of 71.13\%, precision of 78.31\%, and recall of 73.94\%, which were slightly lower than the numbers reported in LADDER's original evaluation (75.32\%, 79.06\%, and 76.98\%, respectively). 
LADDER's lower performance on our test data compared to its reported values was likely due to a distribution shift. 
LADDER's dataset spans 2015 to 2021, while our data is from May 2023 onward. 
This temporal gap may introduce new patterns, terminologies, or threat vectors that LADDER's model struggles to generalize to, even when retrained on a mix of old and new data.

The performance disparity between LADDER and \tool can be attributed to several factors. First, fine-tuning the model in LADDER may lead to overfitting on the training set. Consequently, when confronted with unseen entities in the test set, the model may struggle to recognize them accurately, potentially misclassifying them or recognizing only parts of the entities. 
For example, in the sentence ``... with a specific focus on WordPress sites'', LADDER extracts only ``WordPress'' as an application, resulting in ambiguous content. 
In contrast, \tool correctly extracts ``WordPress sites'', which more accurately reflects the original context. 
Second, similar to EXTRACTOR, the LADDER model lacks sufficient contextual understanding. 
For instance, in the sentence ``The victims include Family Day Care Services, a Canadian childcare service'', LADDER incorrectly identifies ``Canadian'' as a ``B-Location'', whereas it should be recognized as a descriptive term for the childcare service.
Furthermore, LADDER models relation extraction as relation classification, limited to ten relation classes (i.e., \emph{closed-world} setting). 
This constraint restricts the contextual information in the extracted content and hinders the model's ability to generalize to new CTI data containing different or additional relation classes.

\begin{table}[t]
    \centering
    \caption{Performance comparison of \tool and EXTRACTOR on \prefix triplet extraction. }
    \label{tab:EXTRACTOR}
    \centering
\setlength\tabcolsep{3pt}
\scalebox{1}{
    \begin{tabular}{@{}lccc@{}}
        \toprule
        \textbf{Method} & \textbf{F1-Score} & \textbf{Precision} & \textbf{Recall} \\ \midrule
        EXTRACTOR & 62.29 & 51.62
 & 78.53 \\
        \tool & 87.65 & 93.69 & 82.34 \\
        \bottomrule
    \end{tabular}
    }
\end{table}
\begin{table}[t]
    \centering
    \caption{Performance comparison of \tool and LADDER on \prefix entity extraction. }
    \label{tab:LADDER}
    \centering
\setlength\tabcolsep{3pt}
\scalebox{1}{
    \begin{tabular}{@{}lccc@{}}
        \toprule
        \textbf{Method} & \textbf{F1-Score} & \textbf{Precision} & \textbf{Recall} \\ \midrule
 LADDER & 71.13 & 78.31
 & 73.94 \\
        \tool & 90.13 & 92.00 & 88.35 \\
        \bottomrule
    \end{tabular}
    }
    \vspace{-0ex}
\end{table}

\subsection{RQ2: How do different settings affect the cybersecurity triplet extraction?}
\label{subsec:rq2}

To demonstrate the effectiveness of \tool in \prefix triplet extraction, stemming from the superiority of the ICL paradigm and our specific prompt design, we conducted experiments on different ICL configurations, focusing on four aspects: (1) the number of demonstration examples, (2) the permutation of these examples, (3) the backbone model types, and (4) the prompt formulation and design. By default, \tool uses GPT-4 as the model backbone, selects the $k$ most similar 
prompt examples sorting in ascending order of query similarity.

\begin{table}[t]
    \centering
    \caption{Impact of example numbers on \toolno's \prefix triplet extraction.}
    \label{tab:example-num}
    \centering
\setlength\tabcolsep{2pt}
\scalebox{1}{
    \begin{tabular}{@{}ccccc@{}}
        \toprule
        \textbf{Demo. Num.} & \textbf{F1-Score} & \textbf{Precision} & \textbf{Recall} & \textbf{Input\textsubscript{Len}} \\ \midrule
        1 & 85.05 & 94.39 & 77.40 & 949.95\\
        2 & 87.65 & 93.69 & 82.34 & 1539.68\\
        3 & 87.04 & 93.62 & 81.31 & 2138.41 \\
        4 & 86.73 & 89.55 & 84.07 & 2761.38\\ \bottomrule
    \end{tabular}
    }
\end{table}

\paragraph{Impact of prompt example numbers}
\label{subsub:example-num}
To investigate the impact of prompt example numbers, we evaluated 4 configurations: 1, 2, 3, and 4 examples. 
Our observations show effectiveness plateaus when using 2 or 3 examples, while input ICL prompt size increases significantly with more examples. As shown in \cref{tab:example-num}, increasing the prompt example number from 1 to 2 improves all metrics, particularly recall. However, with 3 examples, precision and F1-score plateau, and recall drops by 1\%. 
With 4 examples, recall improves from 82 to 84\%, but precision drops from 93 to 89\%. This contradicts the heuristic that more examples always improve ICL performance but aligns with Chandra et al.~\cite{chandra2024one}, noting that the optimal number of examples varies across scenarios.
Additionally, each additional example increases the input length by an average of 603 tokens, slowing inference speed and increasing computational costs. Thus, our implementation uses two examples in the \prefix triplet extraction phase, balancing effectiveness and efficiency.

\begin{table}[t]
    \centering
    \caption{Impact of example permutation on \toolno's \prefix triplet extraction.} 
    \label{tab:example-permutation}
    \centering
\setlength\tabcolsep{3pt}
\scalebox{1}{
    \begin{tabular}{@{}cccc@{}}
        \toprule
        \textbf{Permutation} & \textbf{F1-Score} & \textbf{Precision} & \textbf{Recall} \\ \midrule
        kNN-ascend & 87.65 & 93.69 & 82.34 \\
        kNN-descend & 85.82 & 90.58 & 81.53 \\
        random & 84.96
 & 90.29 & 80.22 \\ \bottomrule
    \end{tabular}
    }
\end{table}

\paragraph{Impact of prompt example permutations}
\label{subsub:example-perm}
To analyze the effect of the permutation method for selected examples, we examined three strategies: (1) random selection and sorting (random), (2) selection based on kNN similarity and sorting in ascending order (kNN-ascend), and (3) selection based on kNN similarity and sorting in descending order (kNN-descend). These methods were chosen to explore the impact of recency bias in LLMs, which suggests that models give more weight to examples placed nearer to the query~\cite{liu2024lost}. The random method served as a baseline, while kNN-ascend and kNN-descend tested the influence of example order based on similarity. As shown in \cref{tab:example-permutation}, kNN-ascend outperforms other methods across all metrics, indicating the presence of recency bias and its potential for improving results. Consequently, we adopt kNN-ascend for \tool and recommend arranging prompt examples in ascending order of similarity as a universal strategy for other ICL applications.

\paragraph{Impact of backbone models}
\label{para:backbone}
\begin{table}[t]
    \centering

    \caption{Impact of backbone models on \toolno's \prefix triplet extraction.}
    \label{tab:backbone model}
    \centering
\setlength\tabcolsep{3pt}
\scalebox{1}{
    \begin{tabular}{@{}lccc@{}}
        \toprule
        \textbf{Backbone} & \textbf{F1-Score} & \textbf{Precision} & \textbf{Recall} \\ \midrule
        GPT-4 & 87.65 & 93.69 & 82.34 \\
        GPT-3.5 & 76.97 & 82.37 & 72.24 \\
        Qwen2.5-72B & 78.18 & 80.83 & 75.71\\
        Llama3-70B & 77.85 & 81.74 & 74.32\\
        \bottomrule
    \end{tabular}
    }
\end{table}
The emergence of ICL is closely associated with the substantial parameter counts of LLMs. 
To assess \tool's generalizability across different backbone models, we evaluated its performance on representative closed-source LLMs, GPT-3.5 and GPT-4, and leading open-source LLMs, Llama3 and Qwen2.5.
As shown in~\cref{tab:backbone model}, \tool achieves over a 10\% improvement in both recall and precision when using GPT-4 compared to GPT-3.5-turbo. This underscores the importance of leveraging larger models to fully exploit ICL's potential within \tool's framework. 
For Qwen2.5 and Llama3, due to computational resource limitations, we deployed their 72B and 70B parameter versions, respectively. As shown in~\cref{tab:backbone model}, both Qwen2.5 and Llama3 demonstrate performance generally comparable to GPT-3.5-turbo. Specifically, Qwen2.5 exhibits a 1.4\% higher recall but a 0.9\% lower precision compared to Llama3.
GPT-4 excels in both precision and recall among all evaluated backbones. Therefore, all subsequent experiments employ GPT-4 as the default base model.

\paragraph{Prompt formulation and design}
\label{para:prompt_design}
Adhering to our “one-CTI, one-inference” principle, \tool demonstrates exceptional data efficiency. We compare two instruction strategies for prompt design: a multi-round QA-based approach (illustrated in~\cref{fig:EA} and based on~\cite{action}) and our end-to-end prompting approach. In the multi-round method, the extraction prompt comprises three parts: an \emph{instruction} outlining the extraction task, a \emph{context} containing the CTI report, and a \emph{question} requesting specific entities or relations. Consequently, there are \(O\) entity extraction prompts and \(\frac{N \times (N-1)}{2}\) prompts for relation extraction, where \(O\) denotes the number of ontology entity types and \(N\) denotes the number of extracted entities. In contrast, CTINexus consolidates these steps into a single inference, extracting all entities and relations in one round with a uniform prompt template. 
An evaluation of token usage revealed that end-to-end prompting reduces input and output token consumption by 98.9\% and 97.3\%.

In addition, when designing \tool, we iteratively refined the prompt instructions and identified three critical features that significantly boost performance: (1) constraining both output format and content, (2) simulating a \emph{role-playing} scenario, and (3) placing the instruction text at the beginning of the prompt, before any demonstrations. 
Compared to a baseline vanilla prompt without these designs, our refined prompt improved F1-score, precision, and recall by 20.57\%, 15.28\%, and 24.73\%, respectively.

\subsection{RQ3: How do different settings affect the entity alignment and relation prediction?}

\subsubsection{Hierarchical Entity Alignment}
\label{eval:entity-alignment}

As described in \cref{sec:design}, for entity alignment, we first apply ICL to perform coarse-grained grouping of entities based on their types. 
We then vectorize entities into high-dimensional embeddings and conduct fine-grained merging based on semantic similarity.
In the following, we present a series of experiments to investigate the impact of different configurations in entity grouping and merging, aiming to identify the optimal combination.

\begin{table}[t]
    \centering
    \caption{Impact of example numbers on \toolno's coarse-grained entity grouping.}
    \label{tab:ET}
    \centering
\setlength\tabcolsep{3pt}
\scalebox{1}{
    \begin{tabular}{@{}lccc@{}}
        \toprule
        \textbf{Model Config.} & \textbf{Acc.} & \textbf{Micro-F1} & \textbf{Macro-F1} \\ \midrule
        GPT-3.5 (1-shot) & 61.50 & 74.71 & 78.50 \\
        GPT-3.5 (4-shot) & 66.18
 & 78.45 & 79.86 \\
        GPT-3.5 (8-shot) & 69.52 & 80.99 & 82.16 \\ 
        GPT-3.5 (12-shot) & 69.68 & 81.11 & 81.95\\
        \midrule
        GPT-4 (1-shot) & 76.98 & 86.27
 & 86.10 \\
        GPT-4 (4-shot) & 81.02 & 88.94 & 87.87 \\
        GPT-4 (8-shot) & 82.58 & 89.94 & 89.24\\ 
        GPT-4 (12-shot) & 81.18 & 89.05 & 88.28\\
        \bottomrule
    \end{tabular}
    }
\end{table}

\paragraph{Impact of demonstration numbers}
\label{para:example-num4EG}
We assessed the impact of demonstration example numbers on ICL through comparative experiments with four example quantities: 1, 4, 8, and 12. Additionally, we evaluated the performance of two LLMs, GPT-4 and GPT-3.5, across different model sizes.
Notably, the performance showed no significant improvement once the number of examples exceeded 12, so these results were excluded from the table. 
Our evaluation methodology uses accuracy, macro-F1, and micro-F1 metrics, consistent with previous text classification studies~\cite{minaee2021deep}. 
The experimental results, shown in~\cref{tab:ET}, indicate that GPT-4 consistently outperformed GPT-3.5 across all demonstration number hierarchies. 
Remarkably, GPT-4 with 1 demonstration yields better results than GPT-3.5 with 12 demonstrations. 
Both models show substantial improvements when increasing from one to eight demonstrations, but a saturation trend appears when the number of examples exceeds eight. 
This trend is especially evident in GPT-4, where all three metrics slightly decrease as demonstration numbers increase from eight to twelve.

\begin{table}[t]
    \centering
    \caption{Impact of merging threshold values on \toolno's fine-grained entity merging.}
    \label{tab:Threshold}
    \centering
\setlength\tabcolsep{3pt}
\scalebox{1}{
    \begin{tabular}{@{}ccccc@{}}
        \toprule
        \textbf{Threshold} & \textbf{F1-Score} & \textbf{Precision} & \textbf{Recall}& \textbf{Entitiy\textsubscript{Num}} \\ \midrule
        0.4 & 90.10 & 81.99 & 100 & 13.32\\
        0.5 & 95.18 & 90.80 & 100 & 15.13\\
        0.6 & 99.80 & 99.61 & 100 & 16.62\\ 
        0.7 & 96.29 & 99.58 & 93.21 & 17.50\\ \bottomrule
    \end{tabular}
    }
\end{table}

\begin{table}[t]
    \centering
    \caption{Impact of embedding models on \toolno's fine-grained entity merging.}
    \label{tab:Emb}
    \centering
\setlength\tabcolsep{3pt}
\scalebox{1}{
    \begin{tabular}{@{}lcccc@{}}
        \toprule
        \textbf{Model} & \textbf{F1-Score} & \textbf{Precision} & \textbf{Recall} & \textbf{Entitiy\textsubscript{Num}}\\ \midrule
        SecureBERT & 79.15 & 65.50 & 100 & 8.11\\
        text-embedding-3-small & 98.10 & 97.54 & 98.66 & 16.50 \\
        text-embedding-3-large & 99.80 & 99.61 & 100 & 16.62 \\ \bottomrule
    \end{tabular}
    }
\end{table}

\begin{table}[t]
    \centering
    \caption{Impact of example numbers on \toolno's relation prediction.}
    \label{tab:EL}
\centering
\setlength\tabcolsep{3pt}
\scalebox{1}{
    \begin{tabular}{@{}lcccc@{}}
        \toprule
        \textbf{Model Config. } & \textbf{F1-Score} & \textbf{Precision} & \textbf{Recall} \\ \midrule
         GPT-3.5 (0-shot) & 65.95 & 51.26 & 92.42 \\
        GPT-3.5 (1-shot) & 70.21 & 55.46 & 95.65 \\
        GPT-3.5 (2-shot) & 76.87 & 63.31 & 97.84 \\ 
        GPT-3.5 (3-shot) & 74.83 & 61.06 & 96.46\\
        \midrule
        GPT-4 (0-shot) & 85.76 & 75.07
 & 100 \\
        GPT-4 (1-shot) & 89.13 & 80.39 & 100 \\
        GPT-4 (2-shot) & 90.99 & 83.47 & 100\\ 
        GPT-4 (3-shot) & 89.00 & 80.11 & 100\\\bottomrule
    \end{tabular}
}
\end{table}

\begin{table*}[t]
    
    \centering
    
\caption{Averaged token cost and time cost per CTI report for each module and the overall CTINexus pipeline, comparing GPT-4 and GPT-3.5.}
     \label{tab:efficiency}
    \resizebox{\textwidth}{!}{
    \begin{tabular}{lccccccccccccccc}
    \toprule
    \textbf{Backbone} 
    & \multicolumn{2}{c}{\textbf{Cybersecurity Triplet Extraction}} 
    & \multicolumn{2}{c}{\textbf{Hierarchical Entity Alignment}} 
    & \multicolumn{2}{c}{\textbf{Long-Distance Relation Prediction}} 
    & \multicolumn{2}{c}{\textbf{Overall \tool Pipeline}} \\

    \cmidrule(lr){2-3} \cmidrule(lr){4-5} \cmidrule(lr){6-7} \cmidrule(lr){8-9} 

    & Token Cost (\$\//CTI) & Time(s\//CTI)
    & Token Cost (\$\//CTI) & Time(s\//CTI)
    & Token Cost (\$\//CTI) & Time(s\//CTI)
    & Token Cost (\$\//CTI) & Time(s\//CTI)\\
    
    \midrule

    GPT-4  
    & 0.0364 & 11.0905 
    & 0.0393 & 26.1590  
    & 0.0728 & 24.2483 
    & 0.1485 & 67.4865 \\
    
    GPT-3.5 
    & 0.0013 & 5.9824 
    & 0.0018 & 10.6606
    & 0.0038 & 9.5013 
    & 0.0069 & 32.1330 \\
    \bottomrule
\end{tabular}

}
\end{table*}

\paragraph{Impact of embedding models and merging threshold}
The entity merging module applies a text embedding model to vectorize candidate entities grouped by the entity grouping module and uses a merging threshold to identify equivalent entities. 
Here, we evaluated different embedding models and merging thresholds. We used OpenAI’s third-generation embedding models, text-embedding-3-small and text-embedding-3-large, which differ in vector size and represent the latest state-of-the-art general-purpose models. In addition, we also compared with SecureBERT\cite{securebert}, a cybersecurity-specific embedding model based on the RoBERTa architecture pre-trained on a large corpus of cybersecurity data. 
We consider merging thresholds of 0.4, 0.5, 0.6, and 0.7. Besides common metrics for entity alignment, we introduce \textit{Num\_ent}, which records the number of entities after alignment.

The experimental results are shown in~\cref{tab:Threshold,tab:Emb}. 
Threshold values of 0.4, 0.5, and 0.6 all achieve a 100\% recall rate, indicating the algorithm's ability to detect all entities that should be merged. 
However, lower thresholds can erroneously merge non-equivalent entities based on the \textit{Num\_ent} and precision metrics. 
The highest precision is observed when the merging threshold is 0.6. 
Increasing the threshold to 0.7 maintains precision but significantly reduces recall, suggesting overly fine granularity that misclassifies equivalent entities as distinct. 
Regarding embedding models, text-embedding-3-large demonstrates the best performance, with text-embedding-3-small showing similar results. 
SecureBERT, despite its high recall, struggles to correctly cluster entities, as reflected in low precision and \textit{Num\_ent} scores. 
This may be due to the smaller size of RoBERTa compared to the text-embedding-3, leading to less accurate entity distinction. 

\subsubsection{Long-Distance Relation Prediction}
\label{subsubsec:LP}
As mentioned in~\cref{sec:design}, we compose ICL prompts to guide the LLM in inferring relations between disconnected sub-graphs using demonstration examples and context. 
We evaluated different ICL settings by varying the number of demonstration examples and the backbone models. 
Additionally, we examined the effectiveness of zero-shot learning, where the LLM inferred relationships of given entities \emph{without} demonstration examples. 
Zero-shot learning results were excluded from previous ICL experiments due to poor performance. 
The better performance in implicit relation inference compared to other tasks in \tool could be that relation prediction aligns more closely with general NLP tasks. Unlike triplet extraction or entity alignment, which require domain-specific knowledge in the cybersecurity context, relation prediction relies more on LLMs' general ability to infer connections between entities based on linguistic cues. This makes relation prediction less dependent on specialized domain knowledge and more aligned with the LLM's general language understanding capabilities.

Experimental results, shown in~\cref{tab:EL}, indicate that GPT-4 outperforms GPT-3.5 in every setting by a large margin, achieving a 100\% recall rate compared to 92\%-96\% for GPT-3.5. The reason for this discrepancy is that GPT-3.5 has a higher tendency to produce hallucinated answers, either by not following the required instructions for the task (e.g., generating relations between entities not present in the queries) or by not adhering to the required format (e.g., generating a string instead of the requested JSON format). Both models show suboptimal performance with zero-shot learning. Increasing the number of demonstration examples from 1 to 2 significantly improves results, but a slight decline is observed with 3-shot examples. This suggests that while some examples can enhance performance, too many examples may introduce additional complexity or noise.

\subsection{RQ4: How well does \tool perform in end-to-end CSKG construction?}

In contrast to RQ2 and RQ3, which evaluate the effectiveness of each \toolno's component using ground truth inputs, this research question aims to assess the \emph{end-to-end} performance of CTINexus, where each component operates on the output of the preceding one.
We evaluated the final CSKGs constructed by \tool in the end-to-end setting by comparing them to the ground truth CSKGs. 
The evaluation used the triplet-level metrics described in~\cref{para:metrics}, where a triplet was considered correct if its subject, relation, and object semantically match those of a gold-standard triplet.
We evaluated CTINexus with its optimal configuration on ten sampled CTI reports from our dataset, and achieved an end-to-end F1-Score of 87.80\%, a precision of 81.82\%, and a recall of 94.74\%.
These findings confirm that \tool exhibits minimal error propagation across phases, ruling out the snowball effect and ensuring robust utility for downstream applications.

\subsection{RQ5: How well does \tool adapt to different CSKG ontologies?}
To evaluate the adaptability of \tool to various CSKG ontologies, we conducted an experiment using the STIX ontology, a widely recognized threat intelligence-sharing standard~\cite{STIX} supported by numerous vendors. 
STIX categorizes entities into STIX Domain Objects (SDOs) and STIX Relationship Objects (SROs) to systematically capture entities and their interrelationships within threat intelligence. 
For our evaluation, we selected 13 SDOs: Campaign, Grouping, Identity, Indicator, Infrastructure, Intrusion Set, Location, Attack Pattern, Malware, Threat Actor, Tool, Vulnerability, and Report. 
Five SDOs, Course of Action, Note, Opinion, Malware Analysis, and Observed Data, were omitted because they are usually paragraph‑length text blocks that bundle multiple entities, whereas CSKG nodes should represent discrete, meta‑level concepts.
We performed an experiment on a small-scale corpus of ten CTI reports (mean length $\approx 1,318$ tokens) from STIXnet~\cite{STIXnet2024}. 
Manual annotation under the STIX ontology yielded 128 gold‑standard triplets. 
On this set, \tool achieved 89.7 \% precision, 81.9 \% recall, and an 85.6 \% F1‑score—only a few points below its MALOnt performance.
This small drop confirms \toolno’s resilience to ontology shifts: the underlying LLM weights remain frozen, and the target ontology is simply injected into the prompt (and mirrored in a handful of ontology‑consistent demonstration examples), enabling the model to adapt on‑the‑fly to new ontologies.

\subsection{RQ6: What is the efficiency of \tool?}
We assessed the average token and time costs of three modules within \tool, using GPT-3.5 and GPT-4 as backbone models. 
As shown in Table~\ref{tab:efficiency}, the average cost per CTI report is \$0.1485 with GPT-4 and only \$0.0069 with GPT-3.5. The maximum per-report cost reaches \$0.1778 (GPT-4) and \$0.0086 (GPT-3.5), while the minimum costs are \$0.0423 and \$0.0021, respectively.
The results indicate that using GPT-4 as the backbone results in token costs 20-30 times higher than those of GPT-3.5. 
Additionally, the time cost of using GPT-4 is approximately twice as high compared to GPT-3.5 for each module and the overall pipeline. 
The ICL-enhanced relation prediction module is the most computationally expensive, requiring multiple inferences for each input CTI. 
In contrast, the \prefix triplet extraction and hierarchical entity alignment modules have similar token costs, approximately half that of the long-distance relation prediction module, as they adhere to the “one input, one inference” principle, making them more economical. 
Specifically, for the hierarchical entity alignment module, the token and time costs are mainly attributed to the coarse-grained entity grouping module. 
The fine-grained entity merging module, which uses the text-embedding-3-large model, incurs minimal costs (\$0.13 per 1M tokens), resulting in the entire experiment costing less than \$0.30. 

\section{Discussion}
\label{sec:disc}

\paragraph{Limitations} In \tool, the demonstration examples must be carefully chosen and of high quality, with correct answers and the required prompt format. This ensures that \tool can fully utilize the ICL capability to infer the correct answers from the provided examples.
\toolno's performance can degrade if the demonstration set contains incorrect or misformatted samples. Additionally, although \tool can operate in a data-constrained manner, it still requires a certain amount of labeled data, with a recommended minimum of 100 samples. Data imbalance within the demonstration set also affects \toolno's performance, as an imbalanced label distribution leads to less diverse retrieved examples, increasing the likelihood of biased content generation and reducing overall effectiveness.

\paragraph{Hallucinations in LLMs} LLMs can generate hallucinations, which are plausible yet factually inaccurate outputs~\cite{huang2023survey}. For instance, \tool with GPT-3.5 extracted the incorrect triplet $\langle$``July 2022'', ``threat actors behind FARGO attacks were hijacking'', ``vulnerable Microsoft SQL servers''$\rangle$ instead of $\langle$``vulnerable Microsoft SQL servers'', ``are hijacked by'', ``July 2022''$\rangle$, leading to a misplacement of the subject and object and an incoherent relation. This issue is more prevalent in smaller models like GPT-3.5, LLaMA3-70B, and QWen2.5-72B. While potential solutions include fine-tuning hallucination detection classifiers or using stronger LLMs for verification, we leave these challenges for future work. Our current focus is on CSKG construction under data scarcity, where GPT-4 has demonstrated reliable performance and surpasses existing state-of-the-art CTI extraction approaches by a large margin.

\paragraph{Empowering downstream defenses}
\tool has the potential to empower various defensive applications.
For example, the extracted CTI knowledge can be converted (also via LLMs) into open formats like STIX~\cite{STIX}, and exchanged in platforms like AlienVault OTX~\cite{alienvault-otx}, 
and ingested by intrusion prevention systems~\cite{bro1998system,gao2018saql,bajaber2024p4control} and log analysis frameworks~\cite{Hue}. 
A question-answering system can be developed upon the constructed CSKG using LLM's retrieval-augmented generation~\cite{sun2023think}, to provide grounded answers to threat-related questions.
Cyber threat hunting~\cite{threatraptor, POIROT} and investigation~\cite{gao2018aiql,fang2022back} systems can also be enhanced. For example, the effort required for manually constructing investigation queries can be reduced by using LLMs to synthesize or suggest next steps based on the CSKG and partial user input.
We leave the exploration of these downstream applications for future work.

\section{Related Work}
\label{sec:related}

In \cref{sec:bg}, we discussed CTI knowledge extraction works in detail. Here, we discuss additional related work.

\paragraph{CTI services and platforms}
There exist several services that regularly publish updated CTI feeds.
For example, PhishTank~\cite{phishtank} and OpenPhish~\cite{openphish} focus on phishing URLs.
Abuse.ch~\cite{abusech} focuses on malware samples and botnet C\&C servers.
A key limitation is that they only provide isolated IOC feeds.
There are also several comprehensive platforms that allow users to (1) share CTI data with other members of the community to benefit from the crowd-sourced knowledge, or (2) systematically manage their CTI data.
These systems often provide web interfaces for user exploration and APIs for system integration. 
For example, AlienVault OTX~\cite{alienvault-otx} and IBM X-Force Exchange~\cite{ibmxforce} are company-owned crowd-sourced platforms for sharing and searching threat data like IOCs, malware, and vulnerabilities. MISP~\cite{misp} is an open-source platform for sharing, storing, and correlating IOCs of targeted attacks. OpenCTI~\cite{opencti} is an open-source platform that allows users to structure, store, organize, and visualize their CTI knowledge and observables.
Unlike \toolno's automated approach, these platforms require users to actively participate in the sharing process and \emph{manually} contribute CTI data.

\paragraph{Cybersecurity knowledge bases}
Several comprehensive cybersecurity knowledge bases have been created by the industry.
CVE~\cite{cve} and NVD~\cite{nvd} are two most widely used vulnerability databases.
Several threat encyclopedias exist (Trend Micro~\cite{trendmicro}, Kaspersky~\cite{kaspersky-encyclopedia}, F-Secure~\cite{fsecure-encyclopedia}) for malware and vulnerabilities.
MITRE ATT\&CK~\cite{mitreattack} is a knowledge base for cyber adversary tactics and techniques based on real-world observations.
These knowledge bases are manually created by security experts, and hence their update frequency is typically low.
The scope of \tool differs from these systems.
Nevertheless, since these knowledge bases also contain textual CTI descriptions about malware and vulnerabilities,
\tool can be applied to further structuralize such knowledge.

\paragraph{LLMs for cybersecurity} 
Recent works have explored applying LLMs to cybersecurity challenges. 
PentestGPT~\cite{deng2023pentestgpt} investigates LLM capabilities in penetration testing, revealing that while LLMs can handle fundamental tasks and use testing tools competently, they struggle with context loss and attention issues. 
TitanFuzz~\cite{deng2023large} introduces an innovative approach for fuzzing deep-learning libraries using LLMs. 
It employs a generative LLM for high-quality seed programs and an infilling LLM for mutations, significantly improving API and code coverage, and detects numerous previously unknown bugs.
Recent studies have also explored the use of LLMs in tasks such as vulnerability detection~\cite{lin2025from}, 
patch generation~\cite{wei2023copiloting},
malware detection~\cite{zhou2025srdc},
and phishing and scam detection~\cite{li2024knowphish}. 
Unlike these works, \tool leverages the ICL paradigm of LLMs for comprehensive CTI knowledge extraction and CSKG construction.

\paragraph{Other CTI research}
Several studies have empirically examined various aspects of CTI, including understanding vulnerability reproducibility~\cite{mu2018understanding}, evaluating the quality of CTI feeds in terms of volume, timeliness, and coverage~\cite{jin2024sharing,li2019reading}, and analyzing information inconsistencies~\cite{wunder2023shedding}. These works offer valuable insights into the current state of CTI data.
In contrast to these empirical efforts, \tool focuses on designing an LLM-empowered approach for automated extraction of CTI knowledge from CTI reports. The scope is different. 

\section{Conclusion and Future Work}
\label{sec:conclusion}

We presented \tool, a framework leveraging the ICL of LLMs for efficient and adaptive CTI knowledge extraction and CSKG construction. Unlike existing methods, \tool requires minimal data and parameter tuning and can adapt to various ontologies with minimal data annotation. Extensive evaluations demonstrated \tool's effectiveness in extracting comprehensive knowledge, highlighting its potential to transform CTI analysis into a data-efficient and adaptable paradigm.

The rapid adoption of LLMs in security calls for foundation models that can ingest and refresh large‑scale threat intelligence in near real time. 
Current solutions—continual pre‑training on security corpora and retrieval‑augmented generation—are computationally costly, slow to incorporate new data, and struggle to reconcile heterogeneous sources.  
\tool offers an alternative: \emph{KG‑augmented generation}. 
By continuously extracting, aligning, and integrating information into evolving cybersecurity knowledge graphs, \tool can supply downstream security LLMs with a structured, up‑to‑date memory, enabling accurate, cross‑source reasoning at inference time.

Another direction for future work is the integration of visual analytics into \tool. Visual representations of malicious activities could aid analysts in identifying behavior patterns and relationships, offering additional context for understanding threat evolution. Such tools would enhance interpretability and assist in supporting timely cybersecurity decision-making.

\section*{Acknowledgments}
\label{ack}

We would like to thank the anonymous reviewers for their constructive feedback.
This work is supported in part by the Commonwealth Cyber Initiative (CCI). Any opinions, findings, and conclusions made in this paper are those of the authors and do not necessarily reflect the views of the funding agencies.

\bibliographystyle{IEEEtranS}
\bibliography{ref}

\end{document}